\documentstyle[11pt,aaspp4,epsf]{article}
\def\pp{\par\parshape 2 0truecm 15.5truecm 1truecm 14.5truecm\noindent}
\newcommand{\simgt}{\lower.5ex\hbox{$\; \buildrel > \over \sim \;$}}
\newcommand{\simlt}{\lower.5ex\hbox{$\; \buildrel < \over \sim \;$}}

\begin{document}

\begin{minipage}[c]{4cm}
RESCEU-3/99\\
UTAP-318/98\\
\end{minipage}\\

\title{
 Submillimeter detection
   of the Sunyaev -- Zel'dovich effect \\
 toward the most luminous X-ray cluster at z=0.45
}

\bigskip

\author{
Eiichiro Komatsu,\footnotemark[1]
Tetsu Kitayama,\footnotemark[2]
Yasushi Suto,\footnotemark[2]\footnotemark[3]
Makoto Hattori,\footnotemark[1]\\
Ryohei Kawabe,\footnotemark[4] 
Hiroshi Matsuo,\footnotemark[4]
Sabine Schindler,\footnotemark[5]
and Kohji Yoshikawa\footnotemark[6]
}

\bigskip

\footnotetext[1]{Astronomical Institute, T\^{o}hoku University, Aoba,
Sendai 980-8578, Japan}
\footnotetext[2]{Department of Physics, The University of Tokyo,
Tokyo 113-0033, Japan}
\footnotetext[3]{Research Center for the Early University,
School of Science,
The University of Tokyo, Tokyo 113-0033, Japan}
\footnotetext[4]{Nobeyama Radio Observatory, Minamimaki,
Minamisaku, Nagano 384-1305, Japan}
\footnotetext[5]{Astrophysics Research Institute, 
Liverpool John Moores University,
Byrom Street, Liverpool L3 3AF, UK}
\footnotetext[6]{Department of Astronomy, Kyoto University, Kyoto
606-8502, Japan}

\received{1998 November 18}
\accepted{1999 February 24}

\begin{abstract}
  We report on the detection of the Sunyaev -- Zel'dovich (SZ) signals
  toward the most luminous X-ray cluster RXJ1347-1145 at Nobeyama
  Radio Observatory (21 and 43 GHz) and at James Clerk Maxwell
  Telescope (350 GHz). In particular the latter is the first
  successful detection of the SZ temperature increment in the
  submillimeter band which resolved the profile of a cluster of
  galaxies.  Both the observed spectral dependence and the radial
  profile of the SZ signals are fully consistent with those expected
  from the X-ray observation of the cluster. The combined analysis of
  21GHz and 350GHz data reproduces the temperature and core-radius of
  the cluster determined with the ROSAT and ASCA satellites when we
  adopt the slope of the density profile from the X-ray observations.
  Therefore our present data provide the strongest and most convincing
  case for the detection of the submillimeter SZ signal from the
  cluster, as well as in the Rayleigh -- Jeans regime.  We also
  discuss briefly the cosmological implications of the present
  results.
\end{abstract}


\vfill

\centerline{\sl The Astrophysical Journal (Letters), in press}

\eject

\baselineskip 14pt

\section{Introduction}

The Sunyaev -- Zel'dovich (SZ) effect (Zel'dovich \& Sunyaev 1969;
Sunyaev \& Zel'dovich 1972), a change in the apparent brightness of
the cosmic microwave background toward a cluster of galaxies, provides
important probes for cluster gas properties, the global cosmological
parameters and the peculiar velocity field on large scales (e.g. Silk
\& White 1978; Sunyaev \& Zel'dovich 1980; Rephaeli \& Lahav 1991;
Kobayashi, Sasaki \& Suto 1996; Yoshikawa, Itoh \& Suto 1998;
Birkinshaw 1999).  While the temperature decrement due to the SZ
effect is observed for tens of clusters in the Rayleigh -- Jeans
regime, there is no unambiguous SZ detection in the Wien region (i.e.,
the submillimeter band) where the apparent brightness increases.
Andreani et al.  (1996,1999) and Holzapfel et al. (1997) reported the
detection of the SZ temperature increment of the clusters RXJ0658-5557
and A2163, respectively, at the wavelength $\lambda=1.1\sim 1.2$mm.
Although their total fluxes are consistent with the SZ signals from
the clusters, it is not clear to what extent the obtained signals are
affected by other possible contaminations including the dust in our
Galaxy and submm sources in the cluster field (Smail, Ivison \& Blain
1997; Hughes et al. 1998). This question also applies to a recent
claim of the submm SZ detection (Lamarre et al. 1998) toward A2163
which is solely based on the spectral dependence because their
beamsize is very large ($2'\sim3'$).

This simply implies that the mapping observation of the SZ effect is
essential. At cm wavelengths, more than a dozen of clusters have been
already mapped with interferometers, but this technique is not yet
feasible at submm bands. In the present {\it Letter}, we describe our
successful SZ mapping observation of the X-ray cluster RXJ 1347-1145
($z=0.45$) at 350 GHz (0.85mm) with SCUBA (Submillimetre Common-User
Bolometer Array) on JCMT (James Clerk Maxwell Telescope) as well as the
scanning observations (at 21 and 43 GHz) at Nobeyama Radio Observatory.

\section{Observation of the Sunyaev -- Zel'dovich effect\\
  toward RXJ 1347-1145 at cm, mm and submm bands}

\subsection{the target cluster RXJ 1347-1145}

ROSAT and ASCA satellites revealed that RXJ 1347-1145 at $z=0.45$ is
the brightest X-ray cluster of galaxies observed so far (Schindler et
al. 1997). With the additional ROSAT/HRI data acquired recently, the
total exposure time of the cluster in the X-ray observation of RXJ
1347-1145 is now 36.5 ksec.  We have reanalyzed the new X-ray radial
profile, and found that it is well fitted by the isothermal $\beta$
model:
%%%%%%%%%%%%%%%%%%%%%%%%%%%%%%%%%%%%%%%%%%%%%%%%%%%%%%%%%%%%%%%%%%%%%%%
\begin{equation}
  \label{eq:xintensity}
  I_{\rm X}(\theta) \propto n^2_{\rm e0}\theta_{\rm c} 
         [1+(\theta/\theta_{\rm c})^2]^{1/2-3\beta}
\end{equation}
%%%%%%%%%%%%%%%%%%%%%%%%%%%%%%%%%%%%%%%%%%%%%%%%%%%%%%%%%%%%%%%%%%%%%%%
with the following parameters; the central electron density $n_{\rm e0}=
  (9.3\pm0.4)\times 10^{-2}\ {\rm cm^{-3}}$, the core radius
  $\theta_{\rm c}=8.4''\pm 1.0''$, and $\beta=0.57 \pm 0.02$, where
  quoted errors represent 90\% statistical uncertainties (unless
  otherwise stated, we assume $H_0=50$ km/sec/Mpc and $\Omega_0=1.0$
  with vanishing cosmological constant $\lambda_0=0$). Since the
  corresponding SZ radial profile is given by
%%%%%%%%%%%%%%%%%%%%%%%%%%%%%%%%%%%%%%%%%%%%%%%%%%%%%%%%%%%%%%%%%%%%%%%
\begin{equation}
  \label{eq:szintensity}
  I_{\rm SZ}(\theta) \propto T_{\rm e}n_{\rm e0}\theta_{\rm c} 
         [1+(\theta/\theta_{\rm c})^2]^{1/2-3\beta/2} ,
\end{equation}
%%%%%%%%%%%%%%%%%%%%%%%%%%%%%%%%%%%%%%%%%%%%%%%%%%%%%%%%%%%%%%%%%%%%%%%
the cluster is definitely an ideal SZ target due to its unusually
large central density and high temperature $T_e=9.3^{+1.1}_{-1.0}$ keV
(Schindler et al. 1997).  In addition, its small core radius should
enable us to map the radial profile at 350 GHz within the
field-of-view ($160''$) of the SCUBA, while retaining a reasonable
angular resolution (the beamsize $\sigma_{\rm FWHM}$ of SCUBA is
$15''$).  This is the reason why we selected the cluster RXJ 1347-1145
as our SZ target at 350 GHz in the JCMT/SCUBA, as well as at 21 and 43
GHz in the Nobeyama Radio Observatory (NRO).

\subsection{21 and 43 GHz at Nobeyama 45-m telescope}

\begin{figure}
\begin{center}
    \leavevmode\epsfxsize=14cm \epsfbox{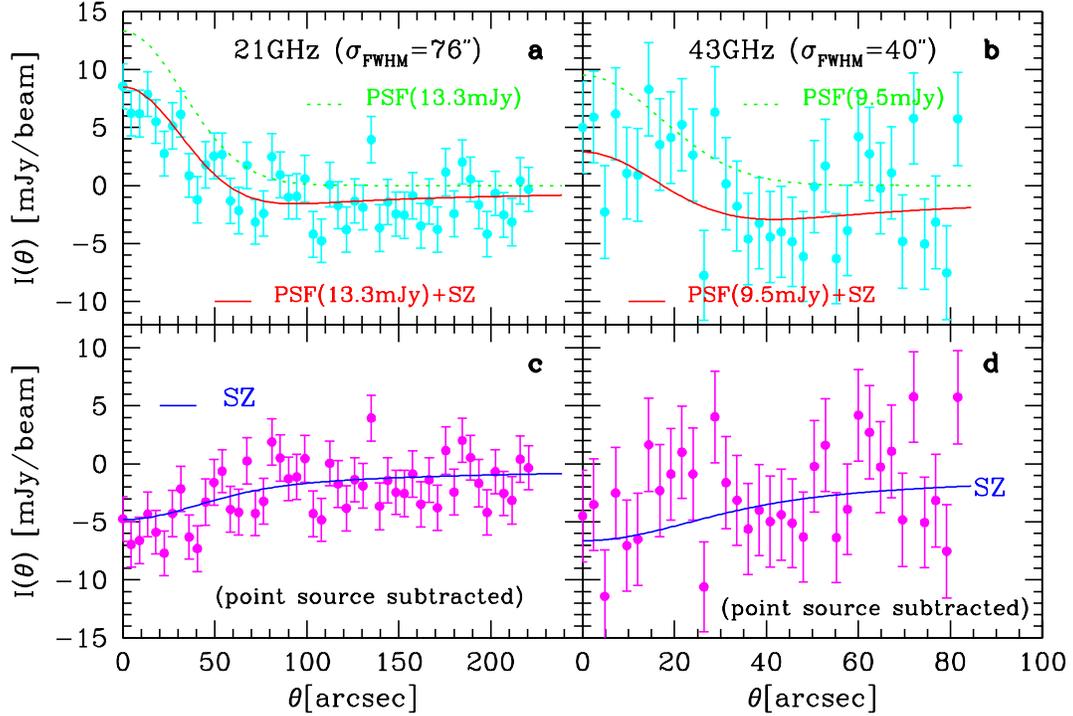}
\caption{
  Radial intensity profile toward RXJ1347 at 21 (left panels) and 43
  (right panels) GHz observed at NRO. Lower panels show the data in
  which the estimated contribution of the central radio source (13.3 mJy
  at 21 GHz, and 9.5 mJy at 43 GHz plotted in green dotted lines) is
  subtracted.  Filled circles indicate our data with $1\sigma$
  error-bars. Red (blue) solid lines in upper (lower) panels are the
  prediction of the signal with (without) the point source
  contribution using the best-fit parameters in the X-ray observation.
\label{fig:2143}
}
\end{center}
\end{figure}

First we observed RXJ 1347-1145 at 21 (43) GHz with HEMT amplifier
(SIS mixer) mounted on the Nobeyama 45-m telescope between March 3rd
and 15th, 1998.  The observation was performed in the cross-scan mode
with $4'\ 45''$ chop throw in azimuth, and NGC7027 was used to
calibrate the flux (estimated calibration error is less than 10\%).  A
total exposure time is 16.2 (14.4) ksec and the beamsize of the PSF
(Point Spread Function) is $\sigma_{\rm FWHM}=76''\ (40'')$, at 21
(43) GHz.  To cut lower frequency noise due to the sky variation, the
data were high-pass filtered with a time-constant of 30 (21) sec, and
then integrated and averaged over radial bins.  The resulting radial
profiles of the cluster at 21 and 43 GHz (upper panels in
Fig.\ref{fig:2143}) indicate the presence of a point source near the
cluster center in addition to the SZ signal. In fact the radio source
was detected also in the NRAO 1.4 GHz sky survey (Condon et al. 1998)
and in the OVRO serendipitous survey of SZ effect at 28.5GHz (Cooray
et al. 1998; Carlstrom, private communication). The former suggests
that the radio source is located at $(\alpha, \delta) = (13^{\rm
  h}47^{\rm m}30.67^{\rm s}, -11^{\circ}\,45'\,8''.6)$ (J2000).  This
is $2.08''$ away from the optical center, $(13^{\rm h}47^{\rm
  m}30.54^{s}, -11^{\circ}\ 45'\,9''.4)$ defined as the location of
the central galaxy.  This $2''$ offset is within a relative positional
error between our radio frame (Johnston et al. 1995) and the optical
frame (MacGillivray \& Stobie 1985) adopted in Schindler et al.
(1995).

% with an
% uncertainty less than $0.5''$, i.e.,
% The position
% uncertainty of the latter is around $1''$. 

\begin{figure}
\begin{center}
    \leavevmode\epsfxsize=10cm \epsfbox{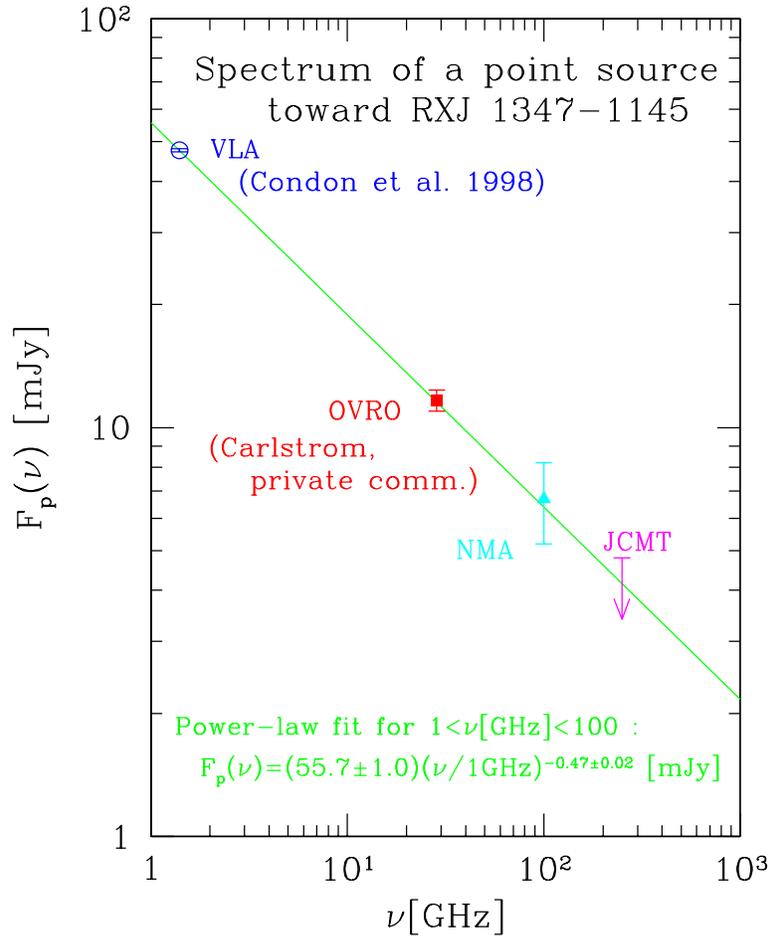}
\caption{
  Spectrum of the central radio source. The data labelled NMA are the
  average of the results at 93 GHz and 105 GHz.
\label{fig:pointflux}
}
\end{center}
\end{figure}

Since the accurate flux of the point source is crucial in properly
extracting the SZ signal, we observed the central source at 93 GHz and
105 GHz simultaneously with Nobeyama Millimeter Array (NMA) between
May 19th and 21st 1998 (15 hours' exposure at each frequency), and at
250 GHz in the photometry mode of SCUBA (Holland et al.  1998) on May
30th and 31st, 1998 (2 hours' exposure). Since the thermal SZ effect
vanishes around at 250 GHz (Rephaeli \& Lahav 1991), the latter
signal, if any, is expected to be dominated by the point source.  We
detected the point source flux of $5.0 \pm 1.5$ mJy at 100 GHz, while
the 250 GHz observation placed a $2\sigma$ upper limit of 4.8 mJy.
These results are summarized in Figure \ref{fig:pointflux}.  We have
corrected the point-source fluxes for the SZ decrement at the
corresponding frequency, although the contamination is comparable or
less than the quoted $1\sigma$ error bars in Figure
\ref{fig:pointflux}.

We fitted the three data of the point source at $\nu\le 100$ GHz to a
single power-law:
%%%%%%%%%%%%%%%%%%%%%%%%%%%%%%%%%%%%%%%%%%%%%%%%%%%%%%%%%%%%%%%%%%%%%
\begin{equation}
  \label{eq:pointflux}
  F_{\rm p}(\nu)
   =(55.7\pm 1.0)(\nu/1{\rm GHz})^{-0.47\pm 0.02} \, {\rm mJy} ,
\end{equation}
%%%%%%%%%%%%%%%%%%%%%%%%%%%%%%%%%%%%%%%%%%%%%%%%%%%%%%%%%%%%%%%%%%%%%
where the quoted errors represent $1\sigma$. Since equation
(\ref{eq:pointflux}) yields a fairly accurate approximation for the
flux at 21 and 43 GHz, we subtract the corresponding contribution of
the point source from our data.  Most radio sources with the spectrum
index less than $\sim 0.5$ are known to exhibit a small amount of time
variation (Eckart, Hummel \& Witzel 1989).  Therefore it is unlikely
that the total flux estimated from equation (\ref{eq:pointflux})
varies significantly due to the possible variability of the source.
The corrected radial profiles of the cluster plotted in lower panels
of Figure \ref{fig:2143} clearly exhibit an extended negative
intensity characteristic of the SZ signal. They are quite consistent
with those expected from the X-ray observation, especially at 21 GHz
where the S/N is significantly higher than at 43 GHz.

\subsection{350 GHz at JCMT/SCUBA}

We observed the cluster at 350 GHz with SCUBA in the jiggling mode on
May 30th and 31st, 1998.  Unfortunately the weather conditions during
our observation were bad (the zenith optical depth at 350 GHz ranged
around $\tau_{350} = 0.46 - 0.60$).  The observation was performed
over 64 independent points over the sky spaced by $3.09''$ each other
with $120''$ chop throw in azimuth.  A total exposure time amounts to
18.6 ksec.  The primary flux calibration and beam measurement were
carried out using Uranus, and the secondary calibrations were
performed at the beginning and the end of each observation using
IRC10216 and 16293-2422, respectively, to check the stability of gain.
The resulting PSF has a beamsize of $\sigma_{\rm FWHM}=15''$, and the
calibration error is less than 15\%. The beam profile was
approximately Gaussian, but our analysis takes account of the effect
of the residual beam-wing as well.

First we analyzed the raw data using REMSKY (Jenness, Lightfoot \&
Holland 1998) in SURF package (Jenness \& Lightfoot 1998) to remove
spatially correlated sky-noise.  With REMSKY we subtract the sky-noise
at each integration from the entire map using the median value of the
bolometers (except for very noisy ones). Therefore the zero-level of
the resulting map after all integrations is still uncertain depending
on the sky condition.  We estimated 1$\sigma$ error of our base-level
or DC offset to be as large as $2.9$ mJy/beam due to the bad weather
conditions.  Then the data were reduced and converted to the image
using SURF ver 1.2.  The resulting image turned out to contain several
bright point sources in the field. We detected 7 spurious
contamination sources above a threshold of 3$\sigma$ using SExtractor
package (Bertin \& Arnouts 1996).  While some of them might simply be
due to the sky noise, others would be real sources; in fact the
previous SCUBA deep surveys toward clusters of galaxies (Smail, Ivison
\& Blain 1997; Smail et al. 1998) and blank fields (Hughes et al.
1998; Barger et al.  1998) detected many submillimeter sources in
their fields (but with $< 10$mJy typically). Since it is premature to
discuss further the reality of the ``sources'' at this point, we
consider the cases with and without the 7 sources, separately.

\begin{figure}
\begin{center}
    \leavevmode\epsfxsize=13cm \epsfbox{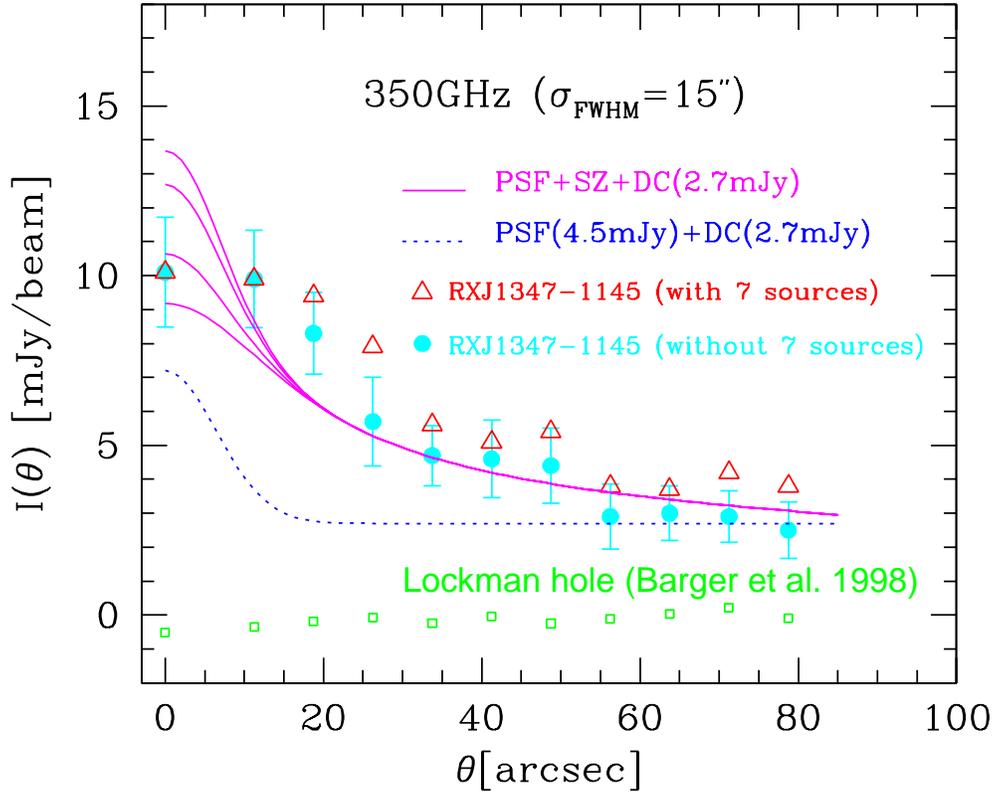}
\caption{
  Radial intensity profile toward RXJ1347 at 350 GHz observed at
  JCMT/SCUBA.  Open triangles (filled circles) indicate our data with
  (without) the 7 spurious sources described in the text. The
  $1\sigma$ error bars are shown only for the latter.  Solid curves
  plot the SZ profiles from the best-fit parameters in the X-ray
  observation and the point-source contribution with $F_{\rm p} = 4.5$
  (a conservative 2$\sigma$ {\it upper} limit from
  eq.[\protect\ref{eq:pointflux}\protect]), $3.5$ (extrapolated from
  eq.[\protect\ref{eq:pointflux}\protect]), $1.5$ (best-fit in Table
  \protect\ref{tab:fit}\protect) and $0$~mJy (from top to bottom). We
  applied the identical reduction procedure to the Lockman-hole data
  (Barger et al. 1998), and the results are plotted in small squares
  for reference (the $1\sigma$ error is smaller than the size of the
  symbol itself).  Dotted curve shows the PSF of 4.5 mJy source with
  2.7 mJy DC offset.
\label{fig:350}
}
\end{center}
\end{figure}

The radially averaged profile of the image is plotted in Figure
\ref{fig:350}. We confirmed that the presence of the extended feature
comparable to the cluster extension is robust even with retaining the
7 sources. In this plot, we adopted the cluster center as the position
of the optical center. It should be noted that the error bars in the
inner annuli are smaller than those in the outer annuli, despite the
smaller effective area.  This is because several noisy bolometers
located at outer annuli contribute significantly to the noise-level of
the corresponding annuli. Thus the noise level is in fact dependent on
the angular radius. These complex noise properties of the SCUBA should
be kept in mind in interpreting the result below.

In fitting our observed profile to the model prediction at 350 GHz, we
have to take account of the DC offset mentioned above, $I_{\rm DC}$,
and the possible contribution of the central point source $F_{\rm p}$.
In what follows, we mainly consider three values for the point-flux so
as to take into account the associated uncertainties; $F_{\rm p}=3.5$
mJy (extrapolated from eq.[\ref{eq:pointflux}]), $4.5$mJy (a
conservative 2$\sigma$ {\it upper} limit from eq.[\ref{eq:pointflux}])
and $0$. We adopt the latter since many radio sources are known to
exhibit a steep decline of flux around submillimeter bands (e.g., Gear
et al. 1994).  Incidentally $F_{\rm p}=4.5$ mJy is a similar flux
level reported for the submm emission from two central galaxies in
luminous X-ray clusters (Edge et al.  1999). Thus the analysis using
$F_{\rm p}=4.5$ mJy may be also useful in understanding the effect of
the possible dust emission from the central galaxy.  Then we treat the
DC offset always as a free fitting parameter. Note that the central
feature of our data is significantly more extended than the PSF of
beam even for the case with $F_{\rm p}=4.5$ mJy ({\it blue dotted
  curve} in Fig.\ref{fig:350}).  The best-fit parameters for $F_{\rm
  p}$ and $I_{\rm DC}$ are listed in Table \ref{tab:fit} together with
the corresponding value of the reduced $\chi^2$.  Table \ref{tab:fit}
indicates that the fit is unacceptable without including the SZ
profile, while the agreement with the SZ profile is insensitive to the
7 sources.

%%%%%%%%%%%%%%%%%%%%%%%%%%%%%%%%%%%%%%%%%%%%%%%%%%%%%%%%%%%%%%%%%%%%%%
\begin{deluxetable}{ccccc}
%\tablenum{1}
\tablewidth{0pt}
\tablecaption{Summary of the best-fit parameters at 350 GHz \label{tab:fit} }
\tablehead{
\colhead{7 sources} &
\colhead{SZ}   & \colhead{$F_{\rm p}$ [mJy]} &
\colhead{$I_{\rm DC}$ [mJy/beam]} & \colhead{reduced $\chi^2$} }
\startdata
No  &  No  & 7.4 & 4.0 & 3.6 \nl
Yes &  No  & 6.1 & 5.1 & 3.6 \nl
No  &  Yes & 1.5 & 2.7 & 0.73 \nl
Yes &  Yes & 0.2 & 3.8 & 0.79 \nl
\enddata
\end{deluxetable}
%%%%%%%%%%%%%%%%%%%%%%%%%%%%%%%%%%%%%%%%%%%%%%%%%%%%%%%%%%%%%%%%%%%%%%

We applied the same reduction procedure for the Lockman-hole data
with SCUBA (Barger et al. 1998), and found no central extended signal
or significant DC offset. This confirms that our signal profile does
not suffer from any systematic effects in the reduction procedure. In
addition, it supports our suspicion that our large DC offset is due to
the relatively large sky-noise during our observing run; our and their
noise levels are typically 8 mJy/beam and 0.8 mJy/beam, respectively.

We repeated the similar fitting analysis in 21 and 43 GHz as well.
The results are summarized in confidence contours on the $T_{\rm e}$ --
$\theta_{\rm c}$ plane (Fig.\ref{fig:thetaT}). Panels a) to c)
indicate that the profile in each band is consistent with each other
and actually in good agreement with the parameters estimated from the
X-ray observation. Combined data analysis of 21 and 350 GHz further
improves the agreement and puts more stringent constraints on $T_{\rm
  e}$ and $\theta_{\rm c}$ (panel d). Therefore our present data
provide the strongest and most convincing case for the detection of
the submm SZ signal from the cluster as well as in the Rayleigh --
Jeans regime.

\section{Discussion}

Detection of the SZ signals in multi-bands for one particular cluster
has important cosmological implications; combining our data in the
X-ray, 21 GHz and 350 GHz of RXJ1347, we estimated the angular
diameter distance at $z=0.451$ as $1897 \pm 317 \pm 246$ Mpc assuming
$F_{\rm p}=3.5$ mJy and the spherical symmetric profile of the cluster
(Silk \& White 1978; Kobayashi, Sasaki \& Suto 1996; Birkinshaw 1999).
The first and second quoted errors come from the uncertainties of the
observed SZ intensity and of the parameters from X-ray observation,
respectively.  This angular diameter distance is translated to $H_0=37
\pm 6 \pm 5$ and $44 \pm 7 \pm 6$ km/sec/Mpc for ($\Omega_0,\lambda_0)
= (1.0,0.0)$ and $(0.3,0.7)$, respectively.  While the estimates still
have fairly large errors compared with those from optical
observations, it is encouraging that they fit in a reasonable range,
and we expect to improve the estimates by observing the cluster again
(hopefully in much better weather condition).  Then it will be
feasible to separate the kinematic and thermal SZ effects by a
simultaneous fit to the 21 and 350 GHz data, which will yield a
estimate of the peculiar velocity of the cluster (Rephaeli \& Lahav
1991; Yoshikawa, Itoh \& Suto 1998; Birkinshaw 1999).

\begin{figure}
\begin{center}
    \leavevmode\epsfxsize=13cm \epsfbox{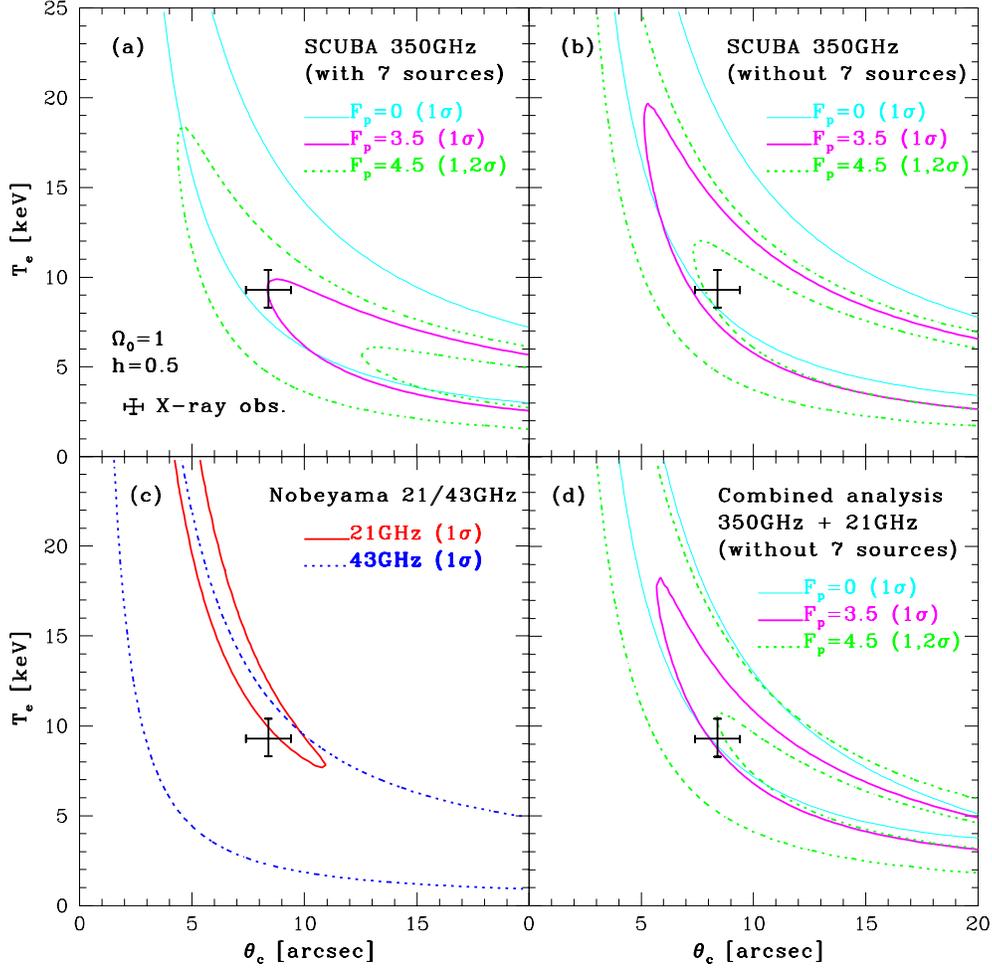}
\end{center}
\caption{
  Confidence contours on the gas temperature $T_e$ and the core radius
  $\theta_{\rm c}$ from the SZ data analysis assuming $\Omega_0=1.0$
  and $h=0.5$ for definiteness. The cross indicates the parameters
  determined from X-ray observations with ROSAT and ASCA satellites.
  a) 350 GHz data with the 7 spurious {\it sources} described in the
  text; b) 350 GHz data without the 7 spurious sources; c) 21 GHz and
  43 GHz; d) combined analysis with 21 GHz and 350 GHz (without the 7
  spurious sources) data. In panels a, b and d, we adopt $F_p=0$mJy
  (thin solid), $F_p=3.5$ mJy (thick solid) and $F_p=4.5$ mJy (thick
  dotted) for the flux of the point source at 350 GHz as explained in
  the text.
\label{fig:thetaT}
}
\end{figure}

In the above discussion, we have neglected several issues which could
affect our interpretation of the detection of the SZ signal from the
cluster in principle, including a possible variability of the central
point source, a non-sphericity and a non-isothermality (Yoshikawa,
Itoh \& Suto 1998; Makino, Sasaki \& Suto 1998; Suto, Sasaki \& Makino
1998; Yoshikawa \& Suto 1999), a cooling flow (Fabian 1994; Allen
1998; Allen \& Fabian 1998), contribution of submm dust (Lamarre et
al. 1998; Edge et al. 1999), unresolved lensed sources, and a peculiar
velocity of the cluster.  These would definitely contribute to put
additional uncertainties in the best-fit parameters of the cluster to
some extent.  Nevertheless, it is almost impossible to explain both
the spectral dependence in three bands and the radial profile in each
band simultaneously by the combination of those effects alone without
the SZ effect, as clearly demonstrated in Figure \ref{fig:thetaT}.
Detailed analysis taking account of those issues will be presented
elsewhere (Komatsu et al., in preparation).

We thank Iain Coulson, Nario Kuno and Satoki Matsushita for kind
assistance during our observing runs at JCMT, Nobeyama 45-m and
Nobeyama Millimeter Array, respectively. We also thank Nick Tothill
for providing the calibration data for our observation, and Tim
Jenness, John Richer, Remo Tilanus and Goeran Sandell for many
fruitful comments and suggestions on data analysis via the SCUBADR
mailing list. We are grateful to A. J. Barger for providing the SCUBA
data of Lockman-Hole, to John Carlstrom for information on the flux of
the central source at 28.5 GHz, and to an anonymous referee for
several critical comments. The travel of E. K. to Hawaii was supported
in part by Satio Hayakawa Foundation in the Astronomical Society of
Japan. T. K. acknowledges support from a JSPS (Japan Society for the
Promotion of Science) fellowship. This research was supported in part
by the Grants-in-Aid for the Center-of-Excellence (COE) Research of
the Ministry of Education, Science, Sports and Culture of Japan to
RESCEU (No.07CE2002).

%%%%%%%%%%%%%%%%%%%%%%%%%%%%%%%%%%%%%%%%%%%%%%%%%%%%%%%%%%%%%%%%%%%   

\newpage
\parskip 0pt
\baselineskip 14pt

\centerline{\bf REFERENCES}

\def\apjpap#1;#2;#3;#4; {\pp#1, {#2}, {#3}, #4}
\def\apjbook#1;#2;#3;#4; {\pp#1, {#2} (#3: #4)}
\def\apjppt#1;#2; {\pp#1, #2}
\def\apjproc#1;#2;#3;#4;#5;#6; {\pp#1, {#2} #3, (#4: #5), #6}

\apjpap Allen, S. W. 1998;MNRAS;296;392;
\apjpap Allen, S. W. \& Fabian, A. C. 1998;MNRAS;297;L57;
\apjpap Andreani, P. et al. 1996;ApJ;459;L49;
\apjppt Andreani, P. et al. 1999;ApJ, in press (astro-ph/9811093);
\apjpap Barger, A. J., Cowie, L. L., Sanders, D. B., Fulton, E.,
Taniguchi, Y., Sato, Y., Kawara, K. \& Okuda, H. 1998;Nature;394;248;
\apjpap Bertin, E. \& Arnouts, S. 1996;A\&AS;117;393;
\apjppt Birkinshaw, M. 1999;Phys. Rep., in press (astro-ph/9808050);
\apjpap Condon, J. J., Cotton, W. D., Greisen, E. W., Yin, Q. F.,
Perley, R. A., Taylor, G. B. \& Broderick, J. J. 1998;AJ;115;1693;
\apjpap Cooray, A. R., Grego, L., Holzapfel, W. L., Joy, M. \&
Carlstrom, J. E. 1998;AJ;115;1388;
\apjpap Eckart, A., Hummel, C. A. \& Witzel, A. 1989;MNRAS;239;381;
\apjppt Edge, A. C., Ivison, R. J. Smail, I., Blain,A.W., \& Kneib, J.-P.
1999;MNRAS, in press (astro-ph/9902038);
\apjpap Fabian, A. C. 1994;ARA\&A;32;277;
\apjpap Gear, W. K., et al. 1994;MNRAS;267;167;
\apjpap Holland, W. S., Cunningham, C. R., Gear, W. K., Jenness, T.,
Laidlaw, K., Lightfoot, J. F. \& Robson, E. I. 1998;Proc. SPIE;3357;305;
\apjpap Holzapfel, W.L., Ade, P.A.R., Church, S.E., Mauskopf, P.D.,
Rephaeli, Y., Wilbanks, T.M., \& Lange, A.E. 1997;ApJ;481;35;
\apjpap Hughes, D. H., et al. 1998;Nature;394;241;
\apjpap Jenness, T., Lightfoot, J. F. \& Holland, W. S. 1998;
Proc. SPIE;3357;548;
\apjppt Jenness T. \& Lightfoot, J. F. 1998;Starlink User Note 216.3;
\apjpap Johnston, K.J. et al. 1995;AJ;110;880;
\apjpap Kobayashi, S., Sasaki, S. \& Suto, Y. 1996;PASJ;48;L107;
\apjpap Lamarre, J. M., et al. 1998;ApJ;507;L5;
\apjpap Makino, N., Sasaki, S. \& Suto, Y. 1998;ApJ;497;555;
\apjpap MacGillivray, H. T. \& Stobie R. S. 1985;Vistas in Astronomy;27;433;
\apjpap Rephaeli, Y. \& Lahav, O. 1991;ApJ;372;21; 
\apjpap Schindler, S. et al. 1995;A\&A;299;L9;
\apjpap Schindler, S., Hattori, M., Neumann, D. M. \& B\"ohringer, H.
1997;A\&A;317;646;
\apjpap Silk, J. \& White, S. D. M. 1978;ApJ;226;L103;
\apjpap Smail, I., Ivison, R. J. \& Blain, A. W. 1997;ApJ;490;L5;
\apjpap Smail, I., Ivison, R. J.,  Blain, A. W. \& Kneib, J.-P. 1998;
ApJ;507;L21;
\apjpap Sunyaev, R. A. \& Zel'dovich, Ya. B. 1972;
Comments Astrophys. Space Phys.;4;173;
\apjpap Sunyaev, R. A. \& Zel'dovich, Ya. B. 1980;MNRAS;190;413;
\apjpap Suto, Y., Sasaki, S. \& Makino, N. 1998;ApJ;509;544;
\apjpap Yoshikawa, K., Itoh, M. \& Suto, Y. 1998;PASJ;50;203;
\apjpap Yoshikawa, K. \& Suto, Y. 1999;ApJ;513;March 10 issue, 
in press (astro-ph/9810247);
\apjpap Zel'dovich, Ya. B. \& Sunyaev, R. A. 1969;
Astrophys. Space. Sci.;4;301;
%%%%%%%%%%%%%%%%%%%%%%%%%%%%%%%%%%%%%%%%%%%%%%%%%%%%%%%%%%%%%%%%%%

\clearpage

%%%%%%%%%%%%%%%%%%%%%%%%%%%%%%%%%%%%%%%%%%%%%%%%%%%%%%%%%%%%%%%%%%%%%%
%%%%%%%%%%%%%%%%%%%%%%%%%%%%%%%%%%%%%%%%%%%%%%%%%%%%%%%%%%%%%%%%%%%%%%

\clearpage

%%%%%%%%%%%%%%%%%%%%%%%%%%%%%%%%%%%%%%%%%%%%%%%%%%%%%%%%%%%%%%%%%%%%%%
%%%%%%%%%%%%%%%%%%%%%%%%%%%%%%%%%%%%%%%%%%%%%%%%%%%%%%%%%%%%%%%%%%%%%%

\clearpage

%%%%%%%%%%%%%%%%%%%%%%%%%%%%%%%%%%%%%%%%%%%%%%%%%%%%%%%%%%%%%%%%%%%%%%
%%%%%%%%%%%%%%%%%%%%%%%%%%%%%%%%%%%%%%%%%%%%%%%%%%%%%%%%%%%%%%%%%%%%%%

\clearpage

%%%%%%%%%%%%%%%%%%%%%%%%%%%%%%%%%%%%%%%%%%%%%%%%%%%%%%%%%%%%%%%%%%%%%%
%%%%%%%%%%%%%%%%%%%%%%%%%%%%%%%%%%%%%%%%%%%%%%%%%%%%%%%%%%%%%%%%%%%%%%

\end{document}